# Chiral determinant on the lattice – Anomalies and Instantons

R. Narayanan[a*] and H. Neuberger[b]

[a]School of Natural Sciences, Institute for Advanced Study, Olden Lane, Princeton, NJ 08540, USA

[b]Department of Physics and Astronomy, Rutgers University, Piscataway, NJ 08855-0849, USA

An expression for the lattice effective action induced by chiral fermions in any even dimensions in terms of an overlap of two states is shown to have promising properties in two and four dimensions: The correct abelian anomaly is reproduced and gauge field configurations with non-zero topological charge are completely suppressed.

## 1. EFFECTIVE ACTION

The lattice effective action induced by chiral fermions, $S_i(U)$, is given by [1,2]

$$e^{S_i(U)} = \frac{{}_U<0-|0+>_U}{{}_1<0-|0+>_1} e^{i[\Phi_+(U)-\Phi_-(U)]}$$

$$e^{i\Phi_\pm(U)} = \frac{{}_U<0\pm|0\pm>_1}{|{}_U<0\pm|0\pm>_1|} \qquad (1)$$

$|0\pm>_U$ are the ground states of the second quantized Hamiltonians $\hat{H}_\pm(U) = -\hat{a}^\dagger \mathbf{H}_\pm(U)\hat{a}$. $\hat{a}^\dagger$ and $\hat{a}$ are fermion creation and annihilation operators. $\mathbf{H}_\pm(U)$ are finite dimensional matrices on a finite lattice that depend on the gauge fields, $U$, and their explicit expressions can be found in [1,2]. $\mathbf{H}_\pm$ have an equal number of positive and negative eigenvalues in the free case. This remains true for perturbative gauge field configurations. It can be proven that $\mathbf{H}_-$ has an equal number of positive and negative eigenvalues for all gauge fields [2]. As such, $S_i(U)$ in (1) is well-defined for a generic gauge field configuration on the lattice. The overlaps of second quantized ground states that appear in (1) are computed by diagonalizing $\mathbf{H}_\pm(U)$. For example, to compute ${}_U<0-|0+>_U$, all the negative eigenvectors, $\Psi_p^\pm$, of $\mathbf{H}_\pm(U)$ are found ($p$ labels the eigenvectors). Let $O_{pq} = \Psi_p^{-\dagger}\Psi_q^+$. Then, ${}_U<0-|0+>_U = \det O$.

*Talk presented at the Lattice '93 conference held between October 12-16 in Dallas, Texas

## 2. U(1) ANOMALY in 2-d and 4-d

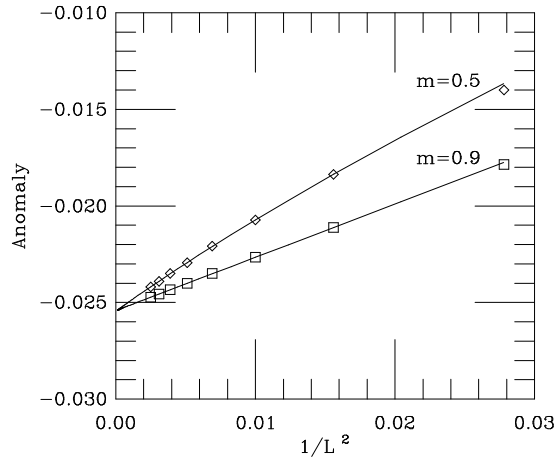

**Figure 5.** The anomaly (LHS of (2)) as a function of lattice size. The lines are fits to the data.

The 2-d U(1) anomaly was computed in [2] as follows. We set $d = 2$ and worked on a square lattice of size $L \times L$. We chose $U_{n,\mu}^{(k)} = \exp[i\frac{A_\mu(\phi)}{L}\cos(\frac{2\pi n \cdot k}{L} + \frac{\pi k_\mu}{L})]$ as the gauge configuration. $k_\mu \in Z$ was kept fixed as $L \to \infty$. $\{U_{n,\mu}^{(k)}\}$ represents a standing plane wave with a fixed number of nodes on the torus in the continuum limit. $A_\mu(\phi) = A_\mu + 2\phi \sin\frac{\pi k_\mu}{L}$ represents a family of gauge equivalent connections. When

$\phi$ is varied, the imaginary part of $S_i(U)$ should change and the change in the continuum limit is

$$\lim_{L\to\infty} -\frac{i}{\pi}\frac{\partial S(A_\mu(\phi))}{\partial\phi}\bigg|_{\phi=0} = \frac{1}{4\pi}[A_1 k_2 - A_2 k_1]. \quad (2)$$

We fixed $k = (1,0)$, $A_1 = A_2 = 0.32$ and, in preparation for $L \to \infty$ limit, computed the LHS in (2) as a function of $L$. The computations were performed at two different values of $m$. The data points are shown in Fig. 1 as a function of $1/L^2$. As expected, the continuum limit is independent of $m$. Extrapolation to $L = \infty$ is smooth and yields $-0.02545(5)$ for the anomaly which agrees with the continuum value of $-0.02546$. Since we used plane waves, once several $k$'s are checked, the numerical work is as good a check as an analytical calculation. We have obtained the correct *consistent* anomaly since the coefficient on the RHS of (2) is $\frac{1}{4\pi}$ and not $\frac{1}{2\pi}$.

The above calculation involves diagonalization of large matrices. In 4-d, gauge fields that produce an anomaly and yet be of the type for which the matrices $e^{\mathbf{H}_\pm(U)}$ are block diagonal of size $O(L^2 \times L^2)$, would simplify the anomaly computation. One such choice is

$$U_{n,4} = e^{i\frac{A_4}{L}\cos\frac{2\pi k_2 n_2}{L}}; \quad U_{n,3} = e^{i\frac{A_3}{L}\cos\frac{2\pi k_1 n_1}{L}}$$

$$U_{n,2} = e^{i\frac{\phi}{L}\left[\cos\frac{2\pi}{L}(k_1 n_1 + k_2 n_2 + k_2) - \cos\frac{2\pi}{L}(k_1 n_1 + k_2 n_2)\right]} \quad (3)$$

$$U_{n,1} = e^{i\frac{\phi}{L}\left[\cos\frac{2\pi}{L}(k_1 n_1 + k_2 n_2 + k_1) - \cos\frac{2\pi}{L}(k_1 n_1 + k_2 n_2)\right]}$$

Since the fields are independent of $n_3$ and $n_4$, the matrices will be block diagonal in momentum space. By varying $\phi$, one generates a family of gauge equivalent configurations. The anomaly equation in the continuum for the above configuration is

$$\lim_{L\to\infty} -\frac{i}{\pi}\frac{\partial S(\phi)}{\partial\phi}\bigg|_{\phi=0} = -\frac{1}{12\pi}A_3 A_4 k_1 k_2 \quad (4)$$

We fixed $k_1 = k_2 = 1$ and $A_3 = A_4 = 0.5$ and computed the LHS of (4) as a function of $L$. The data points are shown in Fig. 2 as a function of $1/L^2$. Extrapolation to $L = \infty$ is smooth and yields $-0.00663(1)$ for the anomaly which agrees with the continuum value of $-0.00663$.

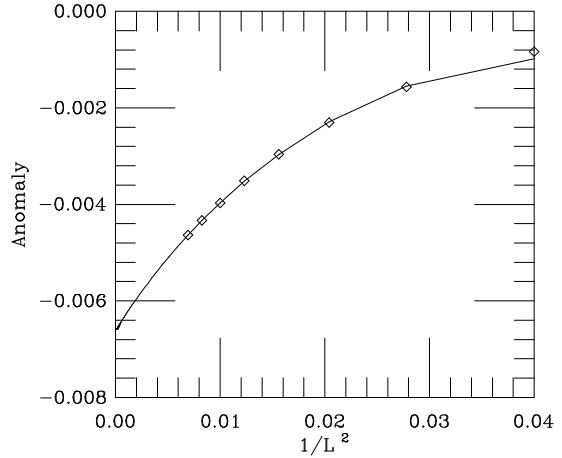

**Figure 2.** The anomaly (LHS of (4)) as a function of lattice size at $m = 0.5$. The line is a fit to the data.

## 3. TOPOLOGICAL CHARGE

The overlap $_U<0-|0+>_U$ will vanish when the numbers of negative and positive eigenvalues of $\mathbf{H}_+(U)$ are different. Insertion of an appropriate number of fermion operators between the ground states will make it non-zero. This is reminiscent of the 't Hooft analysis [3]. Gauge configurations for which this happens have non-zero topological charge and the charge is given by half the difference of the numbers of positive and negative eigenvalues. An example of such a gauge configuration in 2-d is the uniform field $F_{12} = \frac{2\pi q}{L^2}$ for integer $q$. For non-integer values of $q$, the field is not uniform on the lattice but can still give rise to a non-zero topological charge. Since the definition of the topological charge is gauge invariant, any choice of gauge field that generates the above $F_{12}$ is equally good. Fixing $L = 6$, we plot the spectral flow of $\mathbf{H}_+(U)$ as a function of $q$ in Fig. 3. At $L = 6$, $\mathbf{H}_+(U)$ is a $72 \times 72$ matrix and at $q = 0$, it has 36 positive and 36 negative eigenvalues. In between $q = 0$ and $q = 1$, the negative eigenvalue closest to zero crosses zero and becomes positive. Beyond this value of $q$ there is a mismatch; $\mathbf{H}_+(U)$ has 37 positive eigenvalues and 35 negative eigenvalues. The gauge field configurations in this region have a topological charge equal to one. Raising $q$ beyond 1 brings another negative eigenvalue to the positive side making the topological charge equal to two. The situation is reversed when $q$ is negative. The behavior in Fig. 3 is stable against fluctuations of the gauge field.

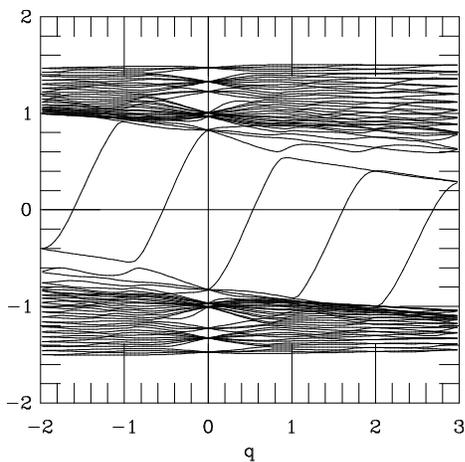

**Figure 3.** Spectral flow of $H_+$ as a function of $q$ at $L = 6$.

To measure the topological charge in 4-d, consider the matrix

$$\mathcal{T}(U) = \frac{1}{2}\left[e^{\mathbf{H}_+(U)} + e^{-\mathbf{H}_+(U)}\right] \quad (5)$$

All eigenvalues of $\mathcal{T}(U)$ are greater than or equal to 1 and every time an eigenvalue of $\mathbf{H}_+(U)$ crosses zero, $\mathcal{T}(U)$ will have an eigenvalue touching 1. Let $q \in [0, 1]$ be some parameter that interpolates between gauge free case ($U(0) = 1$) and the desired gauge configuration ($U(1) = U$). Using a simple power method or Lanczos, we trace the flow of the lowest eigenstate of $\mathcal{T}(U(q))$ as a function of $q$. The number of times the corresponding eigenvalues of $\mathcal{T}(U(q))$ touches one is the topological charge of the gauge configuration. This is same as the number of times the corresponding eigenvalue of $\mathbf{H}_+(U(q))$ crosses zero. If the charge is greater than one in magnitude, there will be discontinuities in the eigenflow of $\mathbf{H}_+(U(q))$ because different eigenstates become the lowest eigenstate of $\mathcal{T}(U(q))$. This method was used in 4-d to study the gauge field configurations that correspond to $F_{12} = F_{34} = \frac{2\pi q}{L^2}$. For integer valued $q$, this configuration has a topological charge equal to $q$. The flow as a function of $q$ is shown in Fig. 4.

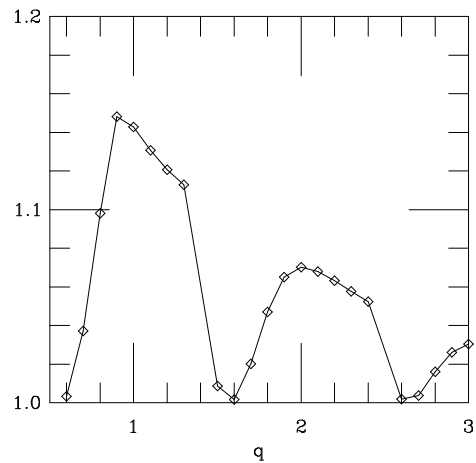

**Figure 4.** Lowest state of $\mathcal{T}(U(q))$ as a function of $q$ at L=8 in 4-d. Diamonds denote the $q$ values where computations were done.

An alternative approach to measure the topological charge is based on the fact that $\mathbf{H}_-(U)$ always has an equal number of positive and negative eigenvalues [2]. There is a natural interpolation between $\mathbf{H}_-(U)$ and $\mathbf{H}_+(U)$ with $m$ being the parameter. The topological charge can be measured by studying the eigenflow of $\mathcal{T}$ as a function of $m$. Details will be provided in the future [4].

This research was supported in part by DOE under grant #DE-FG02-90ER40542 (R.N) and grant #DE-FG05-90ER40559 (H.N).